\documentclass[a4paper,11pt]{article}
\usepackage{pos}
\usepackage{slashed}

\newcommand*{\no}{\noindent}
\newcommand*{\bea}{\begin{eqnarray}}
\newcommand*{\eea}{\end{eqnarray}}
\newcommand*{\be}{\begin{equation}}
\newcommand*{\ee}{\end{equation}}

\newcommand*{\pref}[1]{(\ref{#1})}

\newcommand*{\str}{\mathrm{str}}

\newcommand*{\prefr}[2]{(\ref{#1}-\ref{#2})} 
\newcommand*{\nn}{\nonumber}

\newcommand*{\tr}{\mathrm{tr}}

\newcommand*{\tl}{\mathrm{tl}}

\newcommand{\bma}{\begin{pmatrix}}
\newcommand{\ema}{\end{pmatrix}}

\usepackage{pbox} 
\newcommand{\FPic}[2][{}]{\hspace{-0.27mm}\pbox{\textwidth}{\hspace*{.5ex}\includegraphics[#1]{{#2}}\hspace*{.5ex}}\hspace{-0.27mm}}

\title{The observable spectrum for GUT-like theories}

\author*[a]{Elizabeth Dobson}
\author[a]{Axel Maas}
\author[a,b]{Simon Pl\"atzer}
\author[a,c]{Bernd Riederer}

\affiliation[a]{Institute of Physics, NAWI Graz, Universit\"atsplatz 5, 8010 Graz, Austria}
\affiliation[b]{Particle Physics, Faculty of Physics, University of Vienna, Boltzmanngasse 5, A-1090 Wien, Austria}
\affiliation[c]{BEST - Bioenergy and Sustainable Technologies GmbH, Inffeldgasse 21b, 8010 Graz, Austria}

\emailAdd{elizabeth.dobson@uni-graz.at}
\emailAdd{axel.maas@uni-graz.at}
\emailAdd{simon.plaetzer@uni-graz.at}
\emailAdd{bernd.riederer@uni-graz.at}

\abstract{The spectrum of nonabelian gauge theories cannot be described in terms of elementary particles, and so must be constructed from gauge-invariant composite operators, even in the presence of a Brout–Englert–Higgs effect. This leads to qualitative discrepancies in the prediction of the spectrum between perturbation theory and a full non-perturbative treatment in many theories. This is especially noticeable for GUTs. We present results corroborating this general statement using lattice simulations for a “GUT-like'' toy theory, SU(3) Yang–Mills theory coupled to a Higgs field in the fundamental representation. Despite the apparent simplicity of the model, we find a rich spectrum with some previously unseen features. We also outline the next steps required to generate a large operator basis to extend this investigation to more realistic GUTs.}

\FullConference{The 41st International Symposium on Lattice Field Theory (LATTICE2024)\\
 28 July - 3 August 2024\\
Liverpool, UK\\}

\begin{document}
\maketitle

\section{Introduction}

The observable spectrum of a gauge theory needs to be gauge-invariant. While apparently trivial, this statement has some subtleties in the presence of a Brout–Englert–Higgs (BEH) effect. Since gauge symmetry cannot actually be spontaneously broken due to Elitzur's theorem \cite{Elitzur:1975im}, its breaking in a tree-level treatment \cite{Bohm:2001yx} really just refers to hiding the gauge symmetry via a gauge choice \cite{Maas:2017wzi}. Thus the asymptotic spectrum is still required to be fully manifestly gauge-invariant, not just invariant under perturbative BRST transformations \cite{Maas:2017wzi, Banks:1979fi, Frohlich:1980gj, Frohlich:1981yi}. The failure of perturbative BRST to take care of gauge invariance is due \cite{Fujikawa:1978fu} to the Gribov–Singer ambiguity \cite{Gribov:1977wm,Singer:1978dk}, which is in turn a consequence of the nonabelian structure of the gauge group.

In operative terms, this has surprisingly little consequences for the standard model, which is described by perturbation theory exceedingly well \cite{Bohm:2001yx,pdg}. The reason for this can be traced back to the coincidence of the gauge group and the global symmetry group of the Higgs field, both SU(2). In this very special case the Fr\"ohlich-Morchio-Strocchi (FMS) mechanism \cite{Frohlich:1980gj,Frohlich:1981yi} shows that a one-to-one mapping of the composite spectrum to the elementary spectrum exists. Subleading corrections are present, but are quantitatively small enough to escape detection yet \cite{Maas:2023emb,Maas:2024hcn}. See \cite{Maas:2017wzi,Maas:2023emb} for a review of the standard model case.

This situation changes drastically if the global group is smaller than the gauge group \cite{Maas:2017xzh,Sondenheimer:2019idq}. In this case the gauge-invariant spectrum differs qualitatively from the perturbative (tree-level) spectrum. This has been observed in lattice simulations \cite{Lee:1985yi,Maas:2016ngo,Maas:2018xxu,Afferrante:2020hqe}. However, perturbation theory augmented by the FMS mechanism \cite{Maas:2017wzi,Maas:2024hcn,Dudal:2020uwb,Maas:2020kda} is able to still predict the existing lattice results correctly \cite{Maas:2017xzh}, within uncertainties. This has potentially dramatic implications for BSM model building. Many GUT models are based on differing gauge group and global Higgs group \cite{Bohm:2001yx, Langacker:1980js}. Using FMS-augmented perturbation theory for “realistic" GUTs with minimal Higgs sectors shows that the low-energy spectrum cannot match that of the Standard Model (e.\ g.\ in terms of the number and masses of electroweak vector bosons) \cite{Sondenheimer:2019idq}. Thus, if these predictions are correct, the FMS approach invalidates such GUTs as possible extensions of the standard model.

Unfortunately, realistic GUTs cannot be accessed in lattice simulations, and no experimental hints for them exist, leaving only the FMS results. While this approach has been tested extensively against both lattice simulations and experiment in the standard model, and has always been supported \cite{Maas:2017wzi,Maas:2023emb,Maas:2024hcn,Fernbach:2020tpa}, it is still a paradigmatic shift to move from ordinary perturbation theory to FMS-augmented perturbation theory. This requires as much evidence as possible before relying on it as the only currently known means to perform such predictions reliably. Hence, here we concentrate on further tests using lattice methods in toy GUTs. On the one hand, we extend the previous result for the aforementioned toy GUT \cite{Maas:2016ngo,Maas:2018xxu,Dobson:2022crf} substantially. We find novel, unexpected features in so far unexplored channels. It will be a non-trivial test of FMS-augmented perturbation theory to investigate those. We also outline the necessary steps to extend the existing investigations for a Yang-Mills theory with an adjoint Higgs \cite{Lee:1985yi,Afferrante:2020hqe,Dobson:2022ngz}, which poses substantially new challenges \cite{Dobson:2022ngz} in terms of a suitable operator basis.

In the following we describe the origin of the problem, and its remedy by FMS-augmented perturbation theory, in section \ref{s:theory}. We then discuss the simulations and the spectrum in the fundamental case in section \ref{s:fund}. In the form of an extended outlook we discuss the necessary steps for a comparable analysis with an adjoint Higgs in section \ref{s:adj}. We wrap up in section \ref{s:sum}.

\section{Physical states}\label{s:theory}

Since physical asymptotic states can only be constructed from manifestly gauge-invariant composite  operators, they are classified only in terms of global quantum numbers. We study here a SU(3) Yang-Mills theory coupled to a single scalar Higgs field in either the fundamental representation or the adjoint representation. Thus, besides the usual $J^{PC}$ quantum numbers a state will be characterized by the the global symmetry associated with the Higgs field, which are a U(1) symmetry and a Z$_2$ symmetry, respectively. In perturbation theory, the BEH effect reduces the symmetry to a diagonal subgroup, and the elementary fields are arranged within multiplets of this subgroup \cite{Bohm:2001yx}. A detailed discussion of the perturbative elementary spectrum can be found in, {\it e.g.}, \cite{Maas:2017xzh}. Since the perturbative spectrum is fixed at tree-level, the multiplicities depend on the breaking pattern of the BEH effect. This implies the appearance of symmetry-induced degeneracies, which cannot occur for the global symmetry groups alone.

For the simplest concrete example of where a discrepancy arises, consider our toy GUT with a  fundamental Higgs. At tree-level, we obtain 3 massless gauge bosons, 4 degenerate massive gauge bosons, and a further heavier massive gauge boson \cite{Maas:2017xzh}. For the composite operators and the same $J^{PC}$ assignment this changes, resulting in just a single neutral vector particle \cite{Maas:2017xzh,Maas:2016ngo,Maas:2018xxu}. We furthermore have additional channels, distinguished by their charge under the U(1) custodial symmetry \cite{Maas:2017xzh,Maas:2018xxu}. Since gauge-invariant operators need to carry at least three units of the elementary U(1) charge \cite{Maas:2017xzh}, all such vector states will come in particle-antiparticle pairs with multiples of three times the elementary U(1) charge. Thus, no obvious pattern similar to the perturbative one exists.

The simplest composite vector operators with zero and three times the elementary charge\footnote{The LSZ construction will guarantee that every such operator will source the lightest asymptotic state of these quantum numbers \cite{Bohm:2001yx}. However, in a (FMS-augmented) perturbative expansion this requires to keep the overlap by including (generalized) Bethe-Salpeter amplitudes \cite{Maas:2024hcn,MPS:unpublished}.} are
\bea
O^{1^{--}_0}_i&=&\phi^\dagger D_i\phi\label{c0}\\
O^{1^{--}_1}_i&=&\epsilon_{abc}\phi_a (D^2 \phi)_b (D_i\phi)_c\label{c1},
\eea
\no respectively. FMS-augmented perturbation theory can be used to analyze the content of \prefr{c0}{c1}. The prescription is \cite{Maas:2017wzi,Maas:2023emb,Frohlich:1980gj,Frohlich:1981yi} to fix to a gauge with non-vanishing vacuum expectation value $v$ \cite{Maas:2017wzi} and then to replace the Higgs field $\phi$ by $v+\eta$ with $\eta$ the fluctuation field. Fixing $v$ in the real 3-direction, this yields for the operator \pref{c0} \cite{Maas:2016ngo,Maas:2017xzh}
\be
O^{1^{--}_0}_i=\frac{|v|^2g^2}{4} W_i^8+...\nn
\ee
\no where the other terms contain at least two fields, and thus remain composite. Of course, only the total sum is gauge-invariant. Assuming them to be nonetheless subleading in a suitable sense, it is visible that the operator sources only an object with the same mass as the 8th component of the vector boson. That is the single heaviest one \cite{Maas:2016ngo}. This is indeed the mass which has been observed for the state sourced by \pref{c0} in lattice simulations \cite{Maas:2016ngo,Maas:2018xxu}, supporting this FMS-augmented perturbative result. Hence, the uncharged vector channel is predicted to be qualitatively different from the perturbative prediction.

The situation is more involved for the operator \pref{c1}. Its expansion results in
\bea
O^{1^{--}_1}_i=\frac{|v|^3g^3}{8} s^{abc} W_\mu^a W_\mu^b W_\nu^c+...\nn,
\eea
\no with a non-trivial, but group-theoretically determined, matrix $s$. Thus, it remains to evaluate a six-point function sourced by this composite operator, together with the corresponding Bethe-Salpeter amplitudes, to decide whether a non-trivial asymptotic state is sourced in this way. This is a formidable challenge, and thus guidance by lattice as for the correct answer is invaluable. Especially, as this will decide whether FMS-augmented perturbation theory is sufficient to describe such a state, or whether genuine non-perturbative methods are necessary.

\section{The spectrum with a fundamental Higgs}\label{s:fund}

We therefore provide here a determination of the spectrum of the SU(3) Yang-Mills theory with a fundamental Higgs. Besides the use as a test suite for the methods described in section \ref{s:theory}, it also serves as an example of the possible spectrum of such a theory. Especially, the lightest state with non-zero U(1) charge is necessarily stable. Its existence alone is a unique prediction from gauge-invariance, and its properties therefore an important test.

Technically, the lattice simulations are fairly standard, using the techniques described already in \cite{Dobson:2022ngz}. We investigate channels such as to cover all eighteen continuum spin assignments of spin zero to two, and all combinations of parity and charge parity, for both uncharged and lowest charged channels. We employed a variational analysis, using the techniques of \cite{Jenny:2022atm} with up to 295 operators at rest per continuum channel. We performed simulations along two separate lines connecting a QCD-like domain and a BEH-like domain found in \cite{Maas:2018xxu}. These lines behave as expected for lines of constant physics in each phase, with each having most of the part in either phase. Especially, infinite-volume extrapolated mass ratios tend to be constant within the (in most channels large) statistical errors, implying discretization errors being smaller than statistical uncertainties. We will thus concentrate here on the finest lattice from the long BEH-like line of constant physics, as it is the one most pertaining to the questions raised at the beginning of this section. More details and the other ensembles will be discussed in great detail in an upcoming paper \cite{Dobson:2023}.

As we are ultimately interested in the lowest level in each channel, we will project all lattice irreducible representation to their lowest continuum spin content. Within the variational analysis, different lattice irreducible representations do not mix, as the correlator matrices are block-diagonal. Thus, degeneracies across continuum spins will signal the presence of higher spin contaminations. We do not observe such contaminations for the lowest levels within errors, and thus it is consistent that we do observe the lightest states in the corresponding quantum number channels. However, despite the large operator basis, we do not in all cases observe the lightest scattering state. It is likely that this will require an operator basis including finite-momentum operators \cite{Wurtz:2013ova,Jenny:2022atm}. But this make it still likely that there is no state below the elastic threshold in these channels.

\begin{figure}
\begin{center}
 \includegraphics[width=0.75\textwidth]{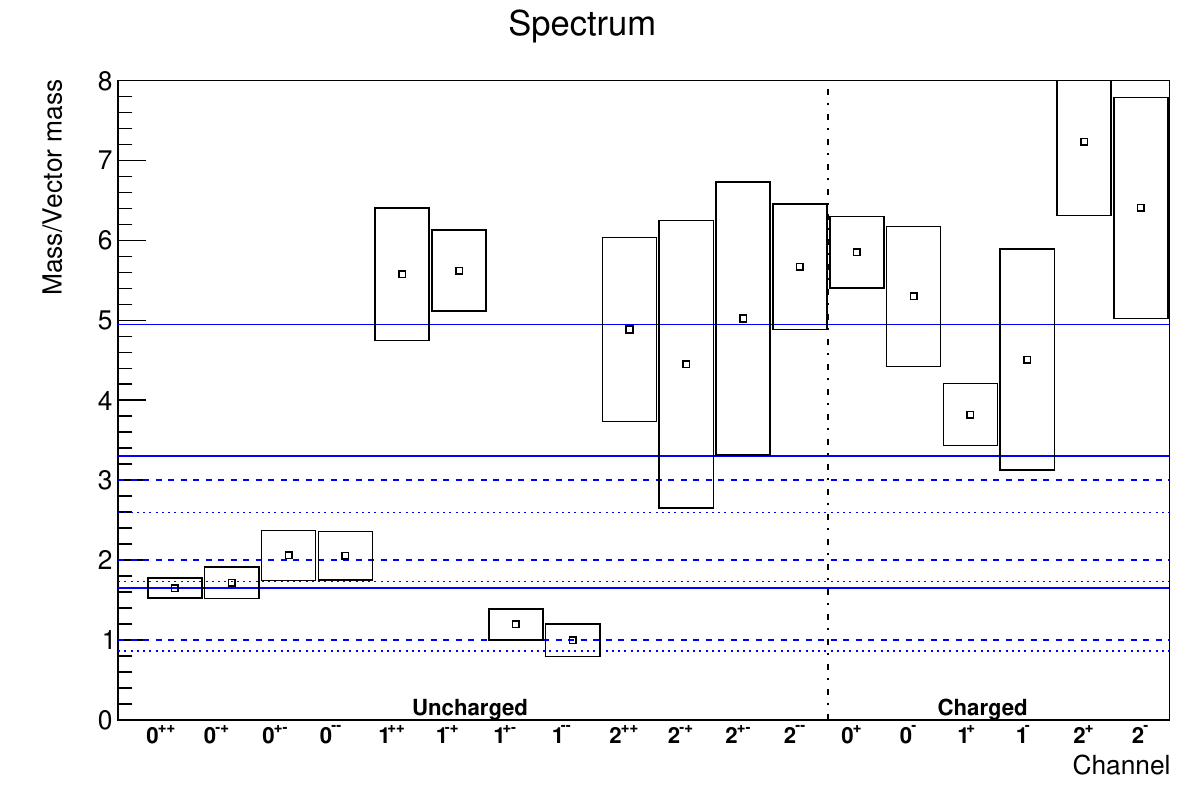}
 \caption{The infinite-volume extrapolated spectrum at $\beta=6.693753$, $\kappa=0.457330$ and $\lambda=3.779690$ \cite{Maas:2018xxu,Dobson:2023} in units of the uncharged vector mass. They are displayed in terms of continuum spin assignments, see text. The horizontal lines are the characteristic tree-level mass scales of the theory, the Higgs mass (full line), the heaviest gauge boson mass (dashed line) and the intermediate gauge boson mass, equal $\sqrt{3/4}$ of the former \cite{Maas:2016ngo}. Charged states carry the smallest possible gauge-invariant U(1) charge, three times the one of the elementary Higgs.}
 \label{spectrum}
 \end{center}
\end{figure}

The results are shown in figure \ref{spectrum}. The vector and the (stable) scalar confirm with the picture of FMS-augmented perturbation theory. The other uncharged states show an interesting pattern. The most striking is that there are (within errors) degenerate states to the scalar and the vector with opposite parity. These quantum number channels were not investigated before, so no prior expectations on them existed. The energy levels are already present without including scattering state operators build with opposite parity operators. They have thus already overlap with non-decomposable operators of these quantum numbers.

A possible explanation for the existence can be motivated from the simplest $1^{+-}$ operator. In the rest frame it takes the form
\be
O^{1^{+-}_0}_i=\epsilon_{ijk}\phi^\dagger D_j D_k\phi=\frac{i|v|^2g^2}{4}s^{ab}\epsilon_{ijk}W_j^a W_k^b+...\label{c2}
\ee
\no To lowest order, this operator is a composite operator of two gauge fields. These can fuse into an $s$-channel gauge boson by a three gauge boson vertex. Analyzing the structure of $s^{ab}$, the resulting terms appear to leading order only to be non-vanishing if the $s$-channel-exchanged gauge boson is indeed the heaviest one, inducing a pole degenerate with the one of the $1^{--}$ state. However, this will require a challenging analysis, also of the Bethe-Salpeter amplitudes, to check whether this ad hoc argument holds up. But this implies there is a motivation born out of FMS-augmented perturbation theory explaining the existence of such a level in this channel. This leads the way to future checks for non-trivial predictions of FMS-augmented perturbation theory using lattice calculations .

Given such an explanation, and assuming a similar one to hold for the scalars, the remainder of the uncharged spectrum is readily explained. The other scalars appear to be consistent with scattering states with (small) relative momenta of two of the vector particles. The other channels are high enough up in the spectrum that they are above the scattering thresholds, thus suggesting no stable states in these channels either.

In the charged sector, a similar degeneracy pattern arises. However, the lightest state, the vector, is substantially heavier than the lightest uncharged one. Thus, it cannot be a simple $s$-channel exchange of an elementary particle to explain either its mass or its degeneracy with its parity partner. The other states are heavy enough that they can be interpreted in terms of a scattering state of the lightest charged state with one of the uncharged states. Thus, the most consistent explanation is that there is only one degenerate pair of stable states, which are the vectors. A very simplistic FMS-augmented constituent model would have placed them at roughly twice the mass of the uncharged vector \cite{Maas:2017xzh}, which appears to be unlikely to really explain the observed masses. Still, it provides the correct ballpark that the charged states should have a mass of a small multiple of the uncharged states. But it also appears unlikely that these states are can be described in a simple constituent model, in which they would have three times the scalar mass. So, their exact nature remains at this time open, and a challenge to an FMS-augmented perturbative analysis.

However, their presence, first hinted at in \cite{Maas:2017xzh,Maas:2018xxu}, seems thereby to be well established. As they cannot exist at all in standard perturbation theory, they add to the differences to such an analysis. Not only are states missing in the uncharged sectors, there are additional charged states, which are substantially heavier. This implies a completely different phenomenology of such theories, which has so far not even be tapped into, {\it e.g.} in terms of hidden sectors.

\section{Towards the spectrum with an adjoint Higgs}\label{s:adj}

In the presence of an adjoint Higgs field, the simulation becomes more expensive due to the presence of massless modes \cite{Lee:1985yi,Afferrante:2020hqe,Dobson:2022ngz}, thus requiring a suitable operator basis to maximize the gain per configuration. Ultimately, the operators used in the fundamental case \cite{Maas:2018xxu,Dobson:2023} do not suffice, and we need to improve it. Especially, in addition to standard pure-gauge operators and traced polynomials of the Higgs fields \cite{Lee:1985yi,Afferrante:2020hqe,Maas:2017xzh}, there are gauge-scalar bound states which have the general form
\begin{equation}
  \mathcal{O}_{\mu_1,\dots,\mu_m; a_1,\dots,a_r}
  =\tr\left[
    (D_{\mu_m} \dots D_{\mu_{a_{r+1}}} \Phi)
    \dots
    (D_{\mu_{a_2}} \dots D_{\mu_{a_1+1}} \Phi)
    (D_{\mu_{a_1}} \dots D_{\mu_1} \Phi)
  \right]
\end{equation}
where $\{a_1,\dots a_r\}$ is some partition of the indices $(\mu_1\dots\mu_m)$. That is, the operators of interest are traces of arbitrary Wilson loops with or without embedded scalars at various points in the loop. On each timeslice one can rewrite the spatial indices to combinations of $D_3$ and $(D_{1}\pm i D_2)$ in order to directly link the spatial parts to components of a spin one triplet. The discrete counterparts of these operators will be loops of links which now also allow for `spikes' branching out of the loop, and any number of scalar fields on the loop, since we encounter $U_\mu(x) \Phi(x+\hat{\mu}) U_{-\mu}(x+\hat{\mu})\ne 1$. While these operators are also demanding in terms of gauge index traces, we here mainly report on how we address their spin properties, which is richer than in the pure gauge case, due to the presence of the above mentioned `spikes' and possibilities to insert scalar fields along the loop.

This complication has led us to investigate
graphical methods to aid the construction of operators, as well as
their subduction to irreducible representations of the octahedral
group: in essence we use birdtrack `completeness relations' \cite{Cvitanovic:2008} at the
level of tensors with vector indices and can either project shapes on
the lattice onto an octahedral irrep $\Lambda$, rather than projecting them
onto an irrep in the continuum (which is similar in spirit of
\cite{Berg:1982kp}), or we can use a continuum operator projected on definite
spin, and subduce it to the octahedral representation, following the
reasoning in \cite{Basak:2005ir,Dudek:2010wm}. Both methods become transparent to be either
side of the same underlying principle by applying graphical tensor
calculus. This will also enable us to formulate rules for the
subduction coefficients. However, it might be algorithmically more
beneficial to start with a set of shapes on the lattice and work our
way towards the continuum representation. In the diagrams we will
discuss lines denote the identities on the space carrying the irrep
indicated on the line, and vertices denote tensor
products\footnote{Through a special normalization one can relate the
  vertices to Clebsh-Gordan coefficients, though we shall not need
  this concept here and rather express all quantities in 3j and 6j
  symbols which have a direct diagrammatic notion to them \cite{Cvitanovic:2008}}.
Irreps can be projected on by building a projector tree
${\mathbf P}^{J_1,...,J_{n-2}}_J$ towards the desired irrep (an empty
label on a line implies the vector representation):
\begin{equation}
{\mathbf P}^{J_1,...,J_{n-2}}_J =  \FPic[scale=0.5]{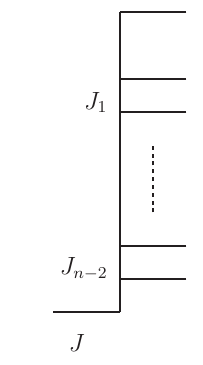}\quad
  \prod_{i=1}^{n-2}\left(\frac{\FPic[scale=0.5]{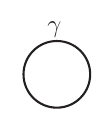}}{\FPic[scale=0.5]{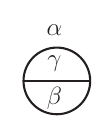}}\right)^{1/2}_{\alpha=1,\beta=J_i,\gamma=J_{i+1}}
\end{equation}
This projector is to be applied to the operator viewed as a tensor
with $n$ legs, one for each fundamental index. The same logic, subject
to the knowledge of the individual vertices and irreps, apply to the
octahedral group, once we formulate the tensor calculus for
${\cal O}_3$ in terms of individual spatial indices ({\it i.e.} just
a rank-$n$ tensor transforming under the discrete group
elements). Different ways of picking those trees can be related to one
another by recoupling relations, and we can choose a default
order without any loss of generality. The completeness and recoupling
relations are well known for ${\rm SU}(2)$ or ${\rm SO}(3)$ but need
to be handled with care for the octahedral group, which we briefly
discuss below.  The subduction coefficients can be obtained
recursively alongside our projector trees from the `radiator' diagram,
schematically as
\begin{equation}
\FPic[scale=0.5]{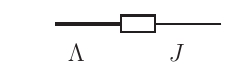}  = \FPic[scale=0.5]{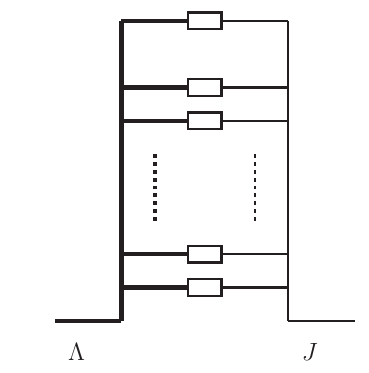} \prod
  \left(\frac{\FPic[scale=0.5]{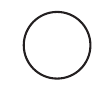}\FPic[scale=0.5]{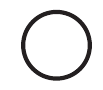}}{
    \FPic[scale=0.5]{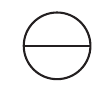}\FPic[scale=0.6]{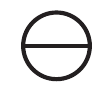}}\right)^{1/2}
\end{equation}
where thick lines now refer to the octahedral group, and boxes denote
subduction coefficients, which are trivial at the level of the vector
representation. In essence this formalism will allow us to construct
tuples of operators and irreps as
$$
(\text{lattice shape}\to \Lambda\to J)\qquad \leftrightarrow\qquad
(\Lambda \leftarrow J \leftarrow \text{continuum operator})
$$
on equal footing and with the same ingredients.

In order to understand how projectors, completeness relations,
recoupling and vertices work on the octahedral group's side, it is
worth noting that, due to its discrete nature, this group admits
additional (actually, an infinite tower of) fundamental invariant
tensors on top of $\delta_{ij}$ and $\epsilon_{ijk}$, namely
$\delta_{i_1,....,i_n}$ for $n\ge 4$, $n$ even, which are defined to
be unity if all indices are equal, and vanish otherwise. We denote
them graphically by an open circle with $n$ legs attached. We need to
account for them in a decomposition of unity, see {\it e.g.}
\cite{Plaetzer:tensordiagrams}, which ultimately leads to the finite
number of irreps ${\cal O}_3$ admits. One finds, {\it e.g.} for
multiplying two fundamental (vector) representations that
\begin{equation}
  {\mathbf P}_{A_1} = \frac{1}{3}\FPic[scale=0.5]{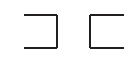}
    \qquad
    {\mathbf P}_{T_1} =\FPic[scale=0.5]{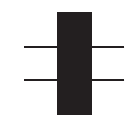}
    \qquad
    {\mathbf P}_{E} =  \FPic[scale=0.5]{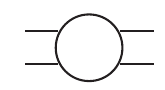} - \frac{1}{3}\FPic[scale=0.5]{singlet}
    \qquad
    {\mathbf P}_{T_2} = \FPic[scale=0.5]{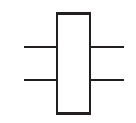} - \FPic[scale=0.5]{delta}
\end{equation}
where we have used the crystallographic labels (black and white boxes
refer to symmetrization and anti-symmetrization), which is to be
compared with the case of ${\rm SO}(3)$, where we have
\begin{equation}
  {\mathbf P}_{J=0} = \frac{1}{3}\FPic[scale=0.5]{singlet}\qquad {\mathbf P}_{J=1} =
  \FPic[scale=0.5]{anti-symmetrizer}\qquad
  {\mathbf P}_{J=2} = \FPic[scale=0.5]{symmetrizer} - \frac{1}{3}\FPic[scale=0.5]{singlet}  .
\end{equation}

Such projectors (acting on tensors representing different ``shapes'' of Wilson loops) allow us to directly find operators in a certain lattice irrep. An advantage of working directly here with the fundamental vector representation is that it makes the subduction from the continuum spin more transparent, which should allow an easier implementation of a large operator basis.

Further constructions along these lines will allow us to construct
more complicated vertices and calculate the underlying 3j, 6j and
subduction coefficients by taking traces of these projectors. This is a demanding combinatorial task for large product spaces. It may be possible that we can derive and solve systems of equations by relating different 3j and 6j symbols with each other along the lines in \cite{Alcock-Zeilinger:2022hrk,Keppeler:2023msu} and follow-up work for ${\rm SU}(N)$. But it will require substantial effort to test this.

\section{Summary}\label{s:sum}

We have provided an extensive analysis of the SU(3) Yang-Mills system
with a fundamental Higgs in the BEH-realization. We find a spectrum
qualitatively different from the one expected in perturbation theory,
despite being at weak coupling \cite{Maas:2016ngo,Maas:2018xxu}. This
implies that even at weak coupling alternative methods are
required. Existing analyses using FMS-augmented perturbation theory
\cite{Maas:2017xzh} are in line with the results, and FMS-mechanism
motivated explanations at least provide some first explanations of
other features of the spectrum. However, especially the charged
sector, will require a very substantial effort to check, whether
FMS-augmented methods are able to describe their features
quantitatively. But given the qualitative failure of the usual
phenomenological tool of perturbation theory for the present case, it
will be necessary to test such methods. Because otherwise no reliable
tools will be available to treat realistic GUTs. On a more technical
side, the demands posed by the adjoint Higgs theory \cite{Lee:1985yi,Afferrante:2020hqe,Dobson:2022ngz} has led us to
develop a novel view on operator constructions on the lattice which we
will present in more detail in a future publication.

\acknowledgments

The computational results presented have been obtained using the Vienna Scientific Cluster (VSC) and the HPC center at the University of Graz. E.\ D.\ and B.\ R.\ have been supported by the Austrian Science Fund FWF, grant P32760.


\begin{thebibliography}{10}

\bibitem{Elitzur:1975im}
S.~Elitzur,
\newblock Phys. Rev. {\bf D12}, 3978 (1975).

\bibitem{Bohm:2001yx}
M.~B\"ohm, A.~Denner, and H.~Joos,
\newblock {\em {Gauge theories of the strong and electroweak interaction}}
  (Teubner, Stuttgart, 2001).

\bibitem{Maas:2017wzi}
A.~Maas,
\newblock Progress in Particle and Nuclear Physics {\bf 106}, 132 (2019),
  1712.04721.

\bibitem{Banks:1979fi}
T.~Banks and E.~Rabinovici,
\newblock Nucl.Phys. {\bf B160}, 349 (1979).

\bibitem{Frohlich:1980gj}
J.~Fr\"ohlich, G.~Morchio, and F.~Strocchi,
\newblock Phys.Lett. {\bf B97}, 249 (1980).

\bibitem{Frohlich:1981yi}
J.~Fr\"ohlich, G.~Morchio, and F.~Strocchi,
\newblock Nucl.Phys. {\bf B190}, 553 (1981).

\bibitem{Fujikawa:1978fu}
K.~Fujikawa,
\newblock Prog. Theor. Phys. {\bf 61}, 627 (1979).

\bibitem{Gribov:1977wm}
V.~N. Gribov,
\newblock Nucl. Phys. {\bf B139}, 1 (1978).

\bibitem{Singer:1978dk}
I.~M. Singer,
\newblock Commun. Math. Phys. {\bf 60}, 7 (1978).

\bibitem{pdg}
Particle Data Group, R.~L. Workman {\em et~al.},
\newblock PTEP {\bf 2022}, 083C01 (2022).

\bibitem{Maas:2023emb}
A.~Maas,
\newblock {\em Rigorous Trails Across Quantum and Classical Physics: A Volume
  in Tribute to Giovanni Morchio} (Springer, 2023), chap. {The
  Fr\"ohlich-Morchio-Strocchi mechanism: A underestimated legacy}, 2305.01960.

\bibitem{Maas:2024hcn}
A.~Maas, D.~M. van Egmond, and S.~Pl\"atzer,
\newblock {Subleading Higgs effects at lepton colliders},
\newblock in {\em {42nd International Conference on High Energy Physics}},
  2024, 2409.20131.

\bibitem{Maas:2017xzh}
A.~Maas, R.~Sondenheimer, and P.~T\"orek,
\newblock Annals of Physics {\bf 402}, 18 (2019), 1709.07477.

\bibitem{Sondenheimer:2019idq}
R.~Sondenheimer,
\newblock Phys. Rev. D {\bf 101}, 056006 (2020), 1912.08680.

\bibitem{Lee:1985yi}
I.-H. Lee and J.~Shigemitsu,
\newblock Nucl. Phys. {\bf B263}, 280 (1986).

\bibitem{Maas:2016ngo}
A.~Maas and P.~T\"orek,
\newblock Phys. Rev. {\bf D95}, 014501 (2017), 1607.05860.

\bibitem{Maas:2018xxu}
A.~Maas and P.~T\"orek,
\newblock Annals Phys. {\bf 397}, 303 (2018), 1804.04453.

\bibitem{Afferrante:2020hqe}
V.~Afferrante, A.~Maas, and P.~T\"orek,
\newblock Phys. Rev. D {\bf 101}, 114506 (2020), 2002.08221.

\bibitem{Dudal:2020uwb}
D.~Dudal {\em et~al.},
\newblock Eur. Phys. J. C {\bf 81}, 222 (2020), 2008.07813.

\bibitem{Maas:2020kda}
A.~Maas and R.~Sondenheimer,
\newblock Phys. Rev. D {\bf 102}, 113001 (2020), 2009.06671.

\bibitem{Langacker:1980js}
P.~Langacker,
\newblock Phys. Rept. {\bf 72}, 185 (1981).

\bibitem{Fernbach:2020tpa}
S.~Fernbach, L.~Lechner, A.~Maas, S.~Pl\"atzer, and R.~Sch\"ofbeck,
\newblock Phys. Rev. D {\bf 101}, 114018 (2020), 2002.01688.

\bibitem{Dobson:2022crf}
E.~Dobson, A.~Maas, and B.~Riederer,
\newblock PoS {\bf LATTICE2022}, 210 (2022), 2211.16937.

\bibitem{Dobson:2022ngz}
E.~Dobson, A.~Maas, and B.~Riederer,
\newblock Annals Phys. {\bf 457}, 169404 (2023), 2211.05812.

\bibitem{MPS:unpublished}
A.~Maas, S.~Pl\"atzer, and R.~Sondenheimer,
\newblock unpublished.

\bibitem{Jenny:2022atm}
P.~Jenny, A.~Maas, and B.~Riederer,
\newblock Phys. Rev. D {\bf 105}, 114513 (2022), 2204.02756.

\bibitem{Dobson:2023}
E.~Dobson, A.~Maas, and B.~Riederer,
\newblock in preparation,
\newblock 2025.

\bibitem{Wurtz:2013ova}
M.~Wurtz and R.~Lewis,
\newblock Phys.Rev. {\bf D88}, 054510 (2013), 1307.1492.

\bibitem{Cvitanovic:2008}
P.~Cvitanovi\'c,
\newblock {\em {Group theory}} (Princeton University Press, Princeton, 2008).

\bibitem{Berg:1982kp}
B.~Berg and A.~Billoire,
\newblock Nucl. Phys. B {\bf 221}, 109 (1983).

\bibitem{Basak:2005ir}
Lattice Hadron Physics (LHPC), S.~Basak {\em et~al.},
\newblock Phys. Rev. D {\bf 72}, 074501 (2005), hep-lat/0508018.

\bibitem{Dudek:2010wm}
J.~J. Dudek, R.~G. Edwards, M.~J. Peardon, D.~G. Richards, and C.~E. Thomas,
\newblock Phys. Rev. D {\bf 82}, 034508 (2010), 1004.4930.

\bibitem{Plaetzer:tensordiagrams}
S.~Pl\"atzer,
\newblock {General tensor diagram calculus},
\newblock unpublished.

\bibitem{Alcock-Zeilinger:2022hrk}
J.~Alcock-Zeilinger, S.~Keppeler, S.~Pl\"atzer, and M.~Sjodahl,
\newblock J. Math. Phys. {\bf 64}, 023504 (2023), 2209.15013.

\bibitem{Keppeler:2023msu}
S.~Keppeler, S.~Pl\"atzer, and M.~Sjodahl,
\newblock JHEP {\bf 05}, 051 (2024), 2312.16688.

\end{thebibliography}

\end{document}